\documentclass[a4paper,fleqn,usenatbib]{mnras}

\usepackage{graphicx}	
\usepackage{amsmath}	
\usepackage{amssymb}	
\usepackage{pdflscape}

\def\msun{\hbox{M$_\odot$}}

\title[Dynamically evolved open clusters]{Highly dynamically evolved intermediate-age open clusters}

\author[Piatti et al.]{
Andr\'es E. Piatti$^{1,2}$\thanks{E-mail: andres@oac.unc.edu.ar}, Wilton S. Dias$^{3}$ and Laura M. Sampedro$^{4,5}$\\
$^{1}$Observatorio Astron\'omico, Universidad Nacional de C\'ordoba, Laprida 854, 5000, 
C\'ordoba, Argentina\\
$^{2}$Consejo Nacional de Investigaciones Cient\'{\i}ficas y T\'ecnicas, Av. Rivadavia 1917, 
C1033AAJ, Buenos Aires, Argentina\\
$^{3}$UNIFEI, Instituto de F\'{\i}sica e Q\'{\i}umica, Universidade Federal de Itajub\'a, Av. BPS 1303 Pinheirinho, 37500-903, Itajub\'a, MG, Brazil\\
$^{4}$Instituto de Astrof\'{\i}sica de Andaluc\'{\i}a, CSIC, Glorieta de
la Astronom\'{\i}a s/n, 18008, Granada, Spain\\
$^{5}$Departamento de Astronomia, IAG,
Universidade de S\~ao Paulo, Rua do Mat\~ao 1226, 
05508-090 S\~ao Paulo, Brazil\\
}

\date{Accepted XXX. Received YYY; in original form ZZZ}

\pubyear{2016}

\begin{document}
\label{firstpage}
\pagerange{\pageref{firstpage}--\pageref{lastpage}}
\maketitle

\begin{abstract}
We present a comprehensive $UBVRI$ and Washington $CT_1T_2$ photometric analysis of 
seven catalogued open clusters, namely: Ruprecht\,3, 9, 37, 74, 150, ESO\,324-15 and 436-2.
The multi-band photometric data sets in combination with 2MASS photometry and {\it Gaia} astrometry for the brighter stars were used to estimate their
structural parameters and fundamental astrophysical properties. We found that Ruprecht\,3
and ESO\,436-2 do not show self-consistent evidence of being physical systems.
The remained studied objects are open clusters of intermediate-age (9.0 $\le$ log($t$ yr$^{-1}$) 
$\le$ 9.6), of relatively small size ($r_{cls}$ $\sim$ 0.4 $-$ 1.3 pc) and placed between 0.6 and 2.9 kpc
from the Sun. We analized the relationships between core, half-mass, tidal and Jacoby radii
as well as half-mass relaxation times to conclude  that the studied clusters are in 
an evolved dynamical stage.
The total cluster masses obtained by summing those of the observed cluster stars
resulted to be $\sim$ 10-15 per cent of the masses of open clusters of similar
age located closer than 2 kpc from the Sun.
We found that cluster stars occupy volumes as large as those for tidally filled clusters.
\end{abstract}




\section{Introduction}

Open clusters evolve dynamically over time due to two-body relaxation and
by the external forces of the interaction with the Galactic tidal field. During this
process open clusters experience structural changes \citep{miholicsetal2014} 
which can be probed from the relationships between core, half-mass and tidal radii \citep{hh03}, 
among others. In this sense, the study of highly dynamically evolved open clusters results 
in a challenging field of research, since they often contain few members. On the other hand, they are suited to test
the particular relationships between their initial masses and dynamical ages.

Identifying dynamically evolved open clusters helps to constrain initial conditions in models
which pursue describing the cluster evolution from N-body simulations \citep{pijlooetal15,rossietal2016},
such as the initial number of stars, the initial mass function, fraction of primordial binaries, etc.
According to the most updated version of the open cluster catalogue compiled by 
\citet[][version 3.5 as of 2016 January]{detal02}, a very limited number of objects have been
studied with some detail from a dynamical point of view \citep{piskunovetal2007}. For this reason,
we have recently started to focus our long-term campaing of improving the statistics of well 
studied open clusters by considering their dynamical evolutions \citep[e.g.][]{piatti16b}.

In this paper, we present for the first time a comprehensive multi-band photometric analysis of 
seven open clusters, namely: Ruprecht\,3, 9, 37, 74, 150, ESO\,324-15 and 436-2 from
 $UBVRI$ and Washington $CT_1T_2$ photometry. Most of them turned out to be open clusters approaching their disruption stage. In Section 2 we describe the collection
and reduction of the available photometric data and their thorough treatment in order to build
extensive and reliable data sets.  The cluster structural (e.g., core and half-light radii) and photometric (e.g., reddening, distance, age, mass) parameters are derived
from star counts and colour-magnitude and colour-colour diagrams, respectively, as described in Sections 3 to 5.
The analysis of the results of the different astrophysical parameters obtained is carried out
in Section 6, where implications are implied.
Finally, Section 7 summarizes the main conclusion of this work.

\section{Data collection and reduction}

We downloaded Johnson $UBV$, Kron-Cousins $RI$ and Washington $C$ images
from the public website of the National Optical Astronomy Observatory 
(NOAO) Science Data Management (SDM) Archives\footnote{http://www.noao.edu/sdm/archives.php.}. 
They were obtained using a 4K$\times$4K CCD 
detector array (scale of 0.289$\arcsec$/pixel) attached to the 1.0-m 
telescope at the Cerro Tololo Inter-American Observatory (CTIO), Chile, in 2011
January 31--February 4  (CTIO program \#2011A-0114, PI: Clari\'a).  
Table~\ref{tab:table1} presents the log of the
observations, where the main astrometric and observational information is 
summarized.

\begin{table*}
\caption{Observations log of selected star clusters.}
\label{tab:table1}
\begin{tabular}{@{}lcccccccc}\hline
Cluster  &R.A.(J2000.0)      &Dec.(J2000.0)     &{\it l} &b       &  filter & exposure & airmass & mean seeing\\
         &(h m s)   &($\degr$ $\arcmin$ $\arcsec$)&(\degr)&(\degr)&  &  (sec)  &  & ($\arcsec$)\\
\hline

Ruprecht\,3& 6 42 6.25&-29 27 18.7& 238.7690 & -14.8182 & $U$ &90, 420& 1.04, 1.03 & 1.2\\
           &          &           &          &          & $B$ &60, 240& 1.02, 1.01 & 1.2\\
           &          &           &          &          & $V$ &20, 120& 1.01, 1.01 & 1.1\\
           &          &           &          &          & $R$ &10, 90& 1.00, 1.00  & 1.1\\
           &          &           &          &          & $I$ &10, 90& 1.00, 1.00  & 1.1\\
           &          &           &          &          & $C$ &90, 360& 1.02, 1.02 & 1.2\\

Ruprecht\,9& 7 2 6.94 &-21 54 40.7& 233.7072 & -7.6073  & $U$ &60, 60, 480& 1.03, 1.02, 1.02 & 1.3\\
           &          &           &          &          & $B$ &20, 60, 360& 1.01, 1.01, 1.01 & 1.2\\
           &          &           &          &          & $V$ &20, 60, 60, 200& 1.02, 1.02, 1.02, 1.01 & 1.2\\
           &          &           &          &          & $R$ &240, 120& 1.01, 1.01 & 1.1\\
           &          &           &          &          & $I$ &10, 90& 1.01, 1.01 & 1.0\\
           &          &           &          &          & $C$ &60, 400& 1.01, 1.01 & 1.2\\

Ruprecht\,37&7 49 46.41&-17 14 53.4& 234.9286 & +4.5499 & $U$ &90, 480& 1.03, 1.03 & 1.2\\
           &          &           &          &          & $B$ &60, 300& 1.04, 1.04 & 1.1\\
           &          &           &          &          & $V$ &20, 60, 180& 1.05, 1.05, 1.05 & 1.0\\
           &          &           &          &          & $R$ &15, 120& 1.06, 1.06 & 1.1\\
           &          &           &          &          & $I$ &10, 90& 1.070, 1.070 & 1.1\\
           &          &           &          &          & $C$ &80, 420& 1.03, 1.03 & 1.1\\

Ruprecht\,74&9 21 0.55& -37 6 42.6 & 263.0370 & +8.9625 & $U$ &60, 240, 540& 1.01, 1.01, 1.01 & 1.4\\
           &          &           &          &          & $B$ &20, 200& 1.03, 1.03 & 1.3\\
           &          &           &          &          & $V$ &15, 180& 1.04, 1.04 & 1.1\\
           &          &           &          &          & $R$ &20, 120& 1.04, 1.04 & 1.1\\
           &          &           &          &          & $I$ &10, 90& 1.05, 1.05 & 1.0\\
           &          &           &          &          & $C$ &120, 480& 1.02, 1.02 & 1.2\\

Ruprecht\,150&7 5 55.22&-28 28 24.4 & 240.0141 &-9.6338 & $U$ &60, 240& 1.01, 1.02 & 1.2\\
           &          &           &          &          & $B$ &20, 180& 1.03, 1.03 & 1.1\\
           &          &           &          &          & $V$ &10, 120& 1.04, 1.04 & 1.1\\
           &          &           &          &          & $R$ &10, 90 & 1.05, 1.05 & 1.0\\
           &          &           &          &          & $I$ &10, 60 & 1.06, 1.06 & 1.1\\
           &          &           &          &          & $C$ &60, 200& 1.02, 1.03 & 1.1\\

ESO\,324-15&13 23 37.3&-41 53 7.5 & 309.3414 & +20.5830 & $U$ &20, 200& 1.04, 1.04 & 1.3\\
           &          &           &          &          & $B$ &10, 150& 1.05, 1.05 &1.1\\
           &          &           &          &          & $V$ &8, 100& 1.06, 1.06 &1.0\\
           &          &           &          &          & $R$ &15, 15& 1.06, 1.07 &1.0\\
           &          &           &          &          & $I$ &10, 80& 1.08, 1.08 & 1.0\\
           &          &           &          &          & $C$ &20, 200& 1.05, 1.05 & 1.3\\

ESO\,436-02&10 14 2.76&-29 11 19.6 & 266.1746 & +22.2431& $U$ &20, 40, 180& 1.01, 1.01, 1.01 & 1.4\\
           &          &           &          &          & $B$ &20, 150& 1.02, 1.03 &1.1\\
           &          &           &          &          & $V$ &4, 8, 100& 1.03, 1.03, 1.03 &1.0\\
           &          &           &          &          & $R$ &3, 60& 1.04, 1.04 &1.0\\
           &          &           &          &          & $I$ &60    & 1.04 &1.0\\
           &          &           &          &          & $C$ &40, 180& 1.02, 1.02 &1.3\\

\hline
\end{tabular}
\end{table*}

All the available series of bias, dome and sky flat exposures per filter during the
observing nights were also downloaded to calibrate the CCD instrumental signature. 
We followed the data reduction procedures documented by the CTIO 
Y4KCam\footnote{http://www.ctio.noao.edu/noao/content/y4kcam} team and utilized the 
{\sc quadred} package in IRAF\footnote{IRAF is distributed by the National 
Optical Astronomy Observatories, which is operated by the Association of 
Universities for Research in Astronomy, Inc., under contract with the National 
Science Foundation.}. Once the calibration frames were properly combined, overscan,
trimming, bias subtraction, flat corrections, etc., were performed.

In order to secure the transformation from the instrumental to the Johnson-Kron-Cousins 
$UBVRI$ and Washington $CT_1T_2$ standard systems, we measured nearly 150 independent 
magnitudes of stars per filter for each night in the standard fields SA\,98 and SA\,101
\citep{l92,g96}, using the {\sc apphot} task within IRAF. The relationships between 
 instrumental (i.e, measured) and standard magnitudes were obtained by fitting the equations:

\begin{equation}
u = u_1 + V + (U-B) + u_2\times X_U + u_3\times (U-B),
\end{equation}

\begin{equation}
b = b_1 + V + (B-V) + b_2\times X_B + b_3\times (B-V),
\end{equation}

\begin{equation}
v = v_1 + V + v_2\times X_V + v_3\times (V-I),			
\end{equation}

\begin{equation}
r = r_1 + V - (V-R) + r_2\times X_R + r_3\times (V-R),
\end{equation}

\begin{equation}
i = i_1 + V - (V-I) + i_2\times X_I + i_3\times (V-I),
\end{equation}

\begin{equation}
c = c_1 + T_1 + (C-T_1) + c_2\times X_C + c_3\times (C-T_1),
\end{equation}

\begin{equation}
r = {t_1}_1 + T_1 + {t_1}_2\times X_{T_1} + {t_1}_3\times (C-T_1),
\end{equation}

\begin{equation}
t_2 = {t_2}_1 + T_1 - (T_1-T_2)  + {t_2}_2\times X_{T_1} + {t_2}_3\times (T_1-T_2),
\end{equation}

\noindent where $u_i$, $b_i$, $v_i$, $r_i$, $i_i$, $c_i$, ${t_1}_i$ and 
${t_2}_i$ ($i$ = 1, 2 and 3) are the fitted coefficients, and $X$ represents the 
effective airmass. Capital and lowercase letters represent standard and instrumental
magnitudes, respectively. Here, we use $r$ magnitudes to derive $T_1$ magnitudes 
because the $R(KC)$ filter is a more efficient substitute of the Washington
$T_1$ filter, as proposed by \citet{g96}. The transformation equations were solved
with the {\sc fitparams} task in IRAF for each night,  and the results are
shown in Table~\ref{tab:table2}.

\begin{figure}
	\includegraphics[width=\columnwidth]{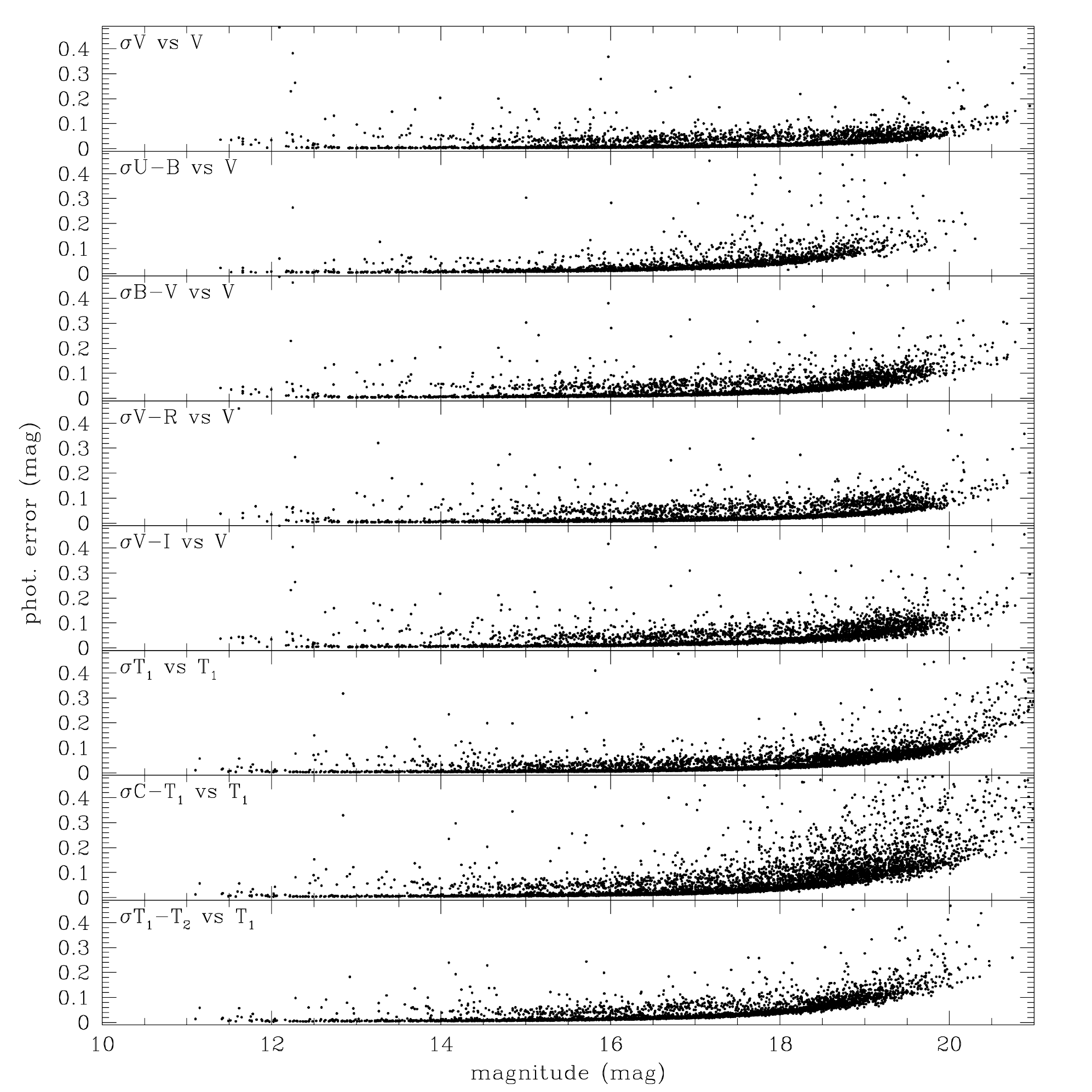}
    \caption{Photometric uncertainties of stars measured in the field of
Ruprecht\,37.}
    \label{fig:fig1}
\end{figure}

Star-finding and point-spread-function (PSF) fitting routines in the {\sc 
daophot/allstar} suite of programs \citep{setal90} were used to derive the stellar
magnitudes. For each image, a quadratically varying PSF was derived by fitting $\sim$ 
200 stars, once the neighbours were eliminated using a preliminary PSF
derived from the brightest, least contaminated $\sim$ 60 stars. We selected both 
groups of PSF stars interactively. We then used the {\sc allstar} program to apply 
the resulting PSF to the identified stars and to create a subtracted image which was 
used to find and measure magnitudes of additional fainter stars. This procedure was 
repeated three times for each frame. After deriving the photometry for all detected 
stars in each filter, a cut was made on the basis of the parameters
returned by DAOPHOT. Fig.~\ref{fig:fig1} illustrates the typical uncertainties in the 
derived photometry. Only objects with $\chi$ $<$2, photometric error less than 2$\sigma$ 
above the mean error at a given magnitude, and $|$SHARP$|$ $<$ 0.5 were kept in each 
image.  Aperture corrections were computed for every measured star and the mean value
for each frame was used.

All individual $u,b,v,r$ and $i$ photometric files were combined into a single master file
using the stand-alone {\sc daomatch} and {\sc daomaster} 
programs\footnote{Kindly provided by P. Stetson.}. We requested that at least one 
colour can be computed during the matching of all the photometric information for each 
star. Similarly, we gathered  the $c,r$, and $i$ photometric files. We thus produced 
2 or 3 independent $ubvri$ and $cri$ data sets, depending on the number of 
observations per filter available. Then we used eqs. 1--8 to standardize the resulting
individual data sets, averaged the standard magnitudes and colours of each star in the 
different data sets, and finally cross-matched the averaged  $UBVRI$ and $CT_1T_2$ data 
sets to build one master table per cluster field. The final information for each cluster
field consists of a running number per star, its $x$ and $y$ coordinates, the mean $V$ 
magnitude, its rms error and the number of measurements, the colours $U-B$, $B-V$,
$V-R$, $V-I$ with their respective rms errors and number of measurements, the $T_1$
magnitude with its error and number of measurements, and the $C-T_1$ and $T_1-T_2$
colours with their respective rms errors and number of measurements.
Table~\ref{tab:table3} gives this information for Ruprech\,3. Only a portion 
of this table is shown here for guidance regarding its form and content. The whole content 
of Table~\ref{tab:table3}, as well as those for the remaining cluster fields
(Tables 4-9), is 
available in the online version of the journal.

\begin{table}
\caption{Mean transformation coefficients for the $UBVRI$ and $CT_1T_2$ photometric systems.}
\label{tab:table2}
\begin{tabular}{@{}ccccc}\hline
Standard     &  zero & extinction & colour& fitting \\
magnitude    &  point   &     coefficient  &  term      & rms     \\\hline

$U$           & 3.296$\pm$0.028  &  0.491$\pm$0.021 &   0.056$\pm$0.022 & 0.071 \\
$B$           & 2.095$\pm$0.014  &  0.327$\pm$0.014 &   0.117$\pm$0.013 & 0.054 \\
$V$           & 1.966$\pm$0.017  &  0.093$\pm$0.012 &  -0.022$\pm$0.010 & 0.050 \\
$R$           & 1.892$\pm$0.014  &  0.095$\pm$0.009 &  -0.003$\pm$0.009 & 0.028 \\
$I$           & 2.828$\pm$0.012  &  0.056$\pm$0.009 &  -0.022$\pm$0.004 & 0.031 \\
$C$           & 1.904$\pm$0.021  &  0.514$\pm$0.017 &  -0.016$\pm$0.012 & 0.043 \\
$T_1$         & 1.911$\pm$0.017  &  0.096$\pm$0.008 &  -0.001$\pm$0.005 & 0.037 \\
$T_2$         & 2.830$\pm$0.023  &  0.045$\pm$0.011 &   0.016$\pm$0.008 & 0.038 \\

\hline
\end{tabular}
\end{table}

\begin{table*}
\caption{$UBVRI$ and $CT_1T_2$ data of stars in the field of Ruprecht\,3.Only a portion 
of this table is shown here for guidance regarding its form and content.}
\label{tab:table3}
\tiny
\begin{tabular}{@{}lcccccccccc}\hline
Star & $x$ & $y$ & $V$ & $U-B$ & $B-V$ & $V-R$ & $V-I$ &
$T_1$ & $C-T1$ & $T_1-T_2$ \\
   & (pixel) & (pixel) & (mag)   & (mag)  & (mag)  & 
(mag)    & (mag)    & (mag)    & (mag)   & 
(mag)   \\\hline
 -- & --& --& -- & --& --& -- & --& --& -- & --  \\
17 & 432.329 & 112.218 &  16.517   0.020  2  &  0.218    0.020  1 &   0.838    0.013  1  &  0.489    0.030  2  &  0.899    0.002  2 &  16.047    0.035  2  &  1.436    0.050  2  &  0.439   0.015  2\\
18 & 820.525 & 125.462  & 16.736    0.014  2  &  0.282    0.026  1  &  0.855    0.016  1  &  0.469    0.016  2  &  0.923    0.052  2 & 16.285    0.013  2  &  1.505    0.011  2  &  0.486    0.055  2\\
19 &2486.025 & 126.433 & 16.068    0.007  2  &  0.119    0.021  2  &  0.701    0.017  2  &  0.446    0.037  2  &  0.893    0.003  2 & 15.641    0.015  2  &  1.199    0.048  2  &  0.479    0.020  2\\
 -- & --& --& -- & --& --& -- & --& --& -- & -- \\
\hline
\end{tabular}

\noindent Columns list a running number per star, its $x$ and $y$ coordinates, 
the mean $V$ 
magnitude, its rms error and the number of measurements, the colours $U-B$, $B-V$,
$V-R$, $V-I$ with their respective rms errors and number of measurements, the $T_1$
magnitude with its error and number of measurements, and the $C-T_1$ and $T_1-T_2$
colours with their respective rms errors and number of measurements.
\end{table*}

\section{Cluster structural parameters}

Stellar density radial profiles were built once we determined the geometrical centres 
of the clusters. In order to do that we fitted Gaussian distributions to the star counts in the $x$ 
and $y$ directions for each cluster. The fits of the Gaussians were performed using the 
{\sc ngaussfit} routine in the {\sc stsdas/iraf} package. We adopted a single Gaussian and 
fixed the constant to the corresponding background levels (i.e. stellar field densities 
assumed to be uniform) and the linear terms to zero. The centre of the Gaussian, its 
amplitude, and its $FWHM$ acted as variables. The number of stars projected along the $x$ 
and $y$ directions were counted within intervals of 20, 40, 60, 80 and 100 pixel wide, and 
the Gaussian fits repeated each time. Finally, we averaged the five different Gaussian 
centres  resulting a typical standard deviation of $\pm$  50 pixels ($\pm$ 14.5$\arcsec$) in  all cases. 

Subsequently stellar density profiles based on star counts previously performed within boxes 
of 50 pixels per side distributed throughout the whole field of each cluster were built. 
The chosen box size allowed us to sample the stellar spatial distribution statistically. 
Thus, the number of stars per unit area at a given radius $r$ can be directly calculated 
through the expression:

\begin{equation}
(n_{r+25} - n_{r-25})/(m_{r+25} - m_{r-25}),
\end{equation}

\noindent where $n_r$ and $m_r$ represent the number of stars and boxes, respectively,  
included in a circle of radius $r$. Note that this method does not necessarily require a 
complete circle of radius $r$ within the observed field to estimate the mean stellar
density at that distance. With a stellar density profile that extends far away from 
the cluster centre -but not too far so as to risk losing the local field-star signature- 
it is possible to estimate the background level with 
high precision, which is particularly useful when dealing with loose clusters.
On the other hand, the more accurate the  background level the more precise the cluster
 radius ($r_{cls}$), defined here as the distance from the cluster centre where 
the observed density profile intersecs the background level (see Table~\ref{tab:table10}). 
 We computed the average and 
corresponding rms error of the background level at any distance to the cluster centre 
by using every available star
count measurement at that distance. Then, the mean background and its error was 
calculated by averaging all these latter values.

\begin{figure*}
	\includegraphics[width=\columnwidth]{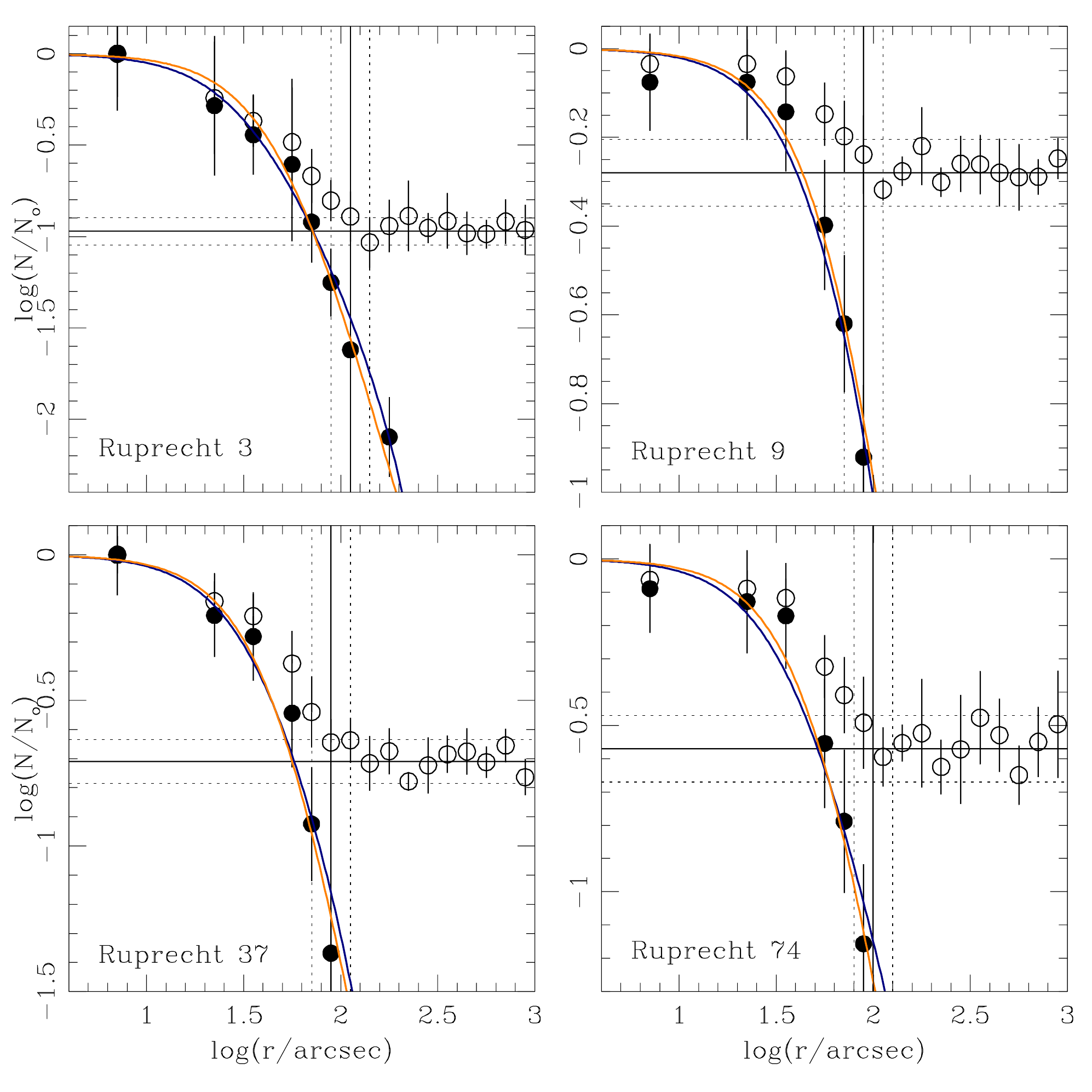}
        \includegraphics[width=\columnwidth]{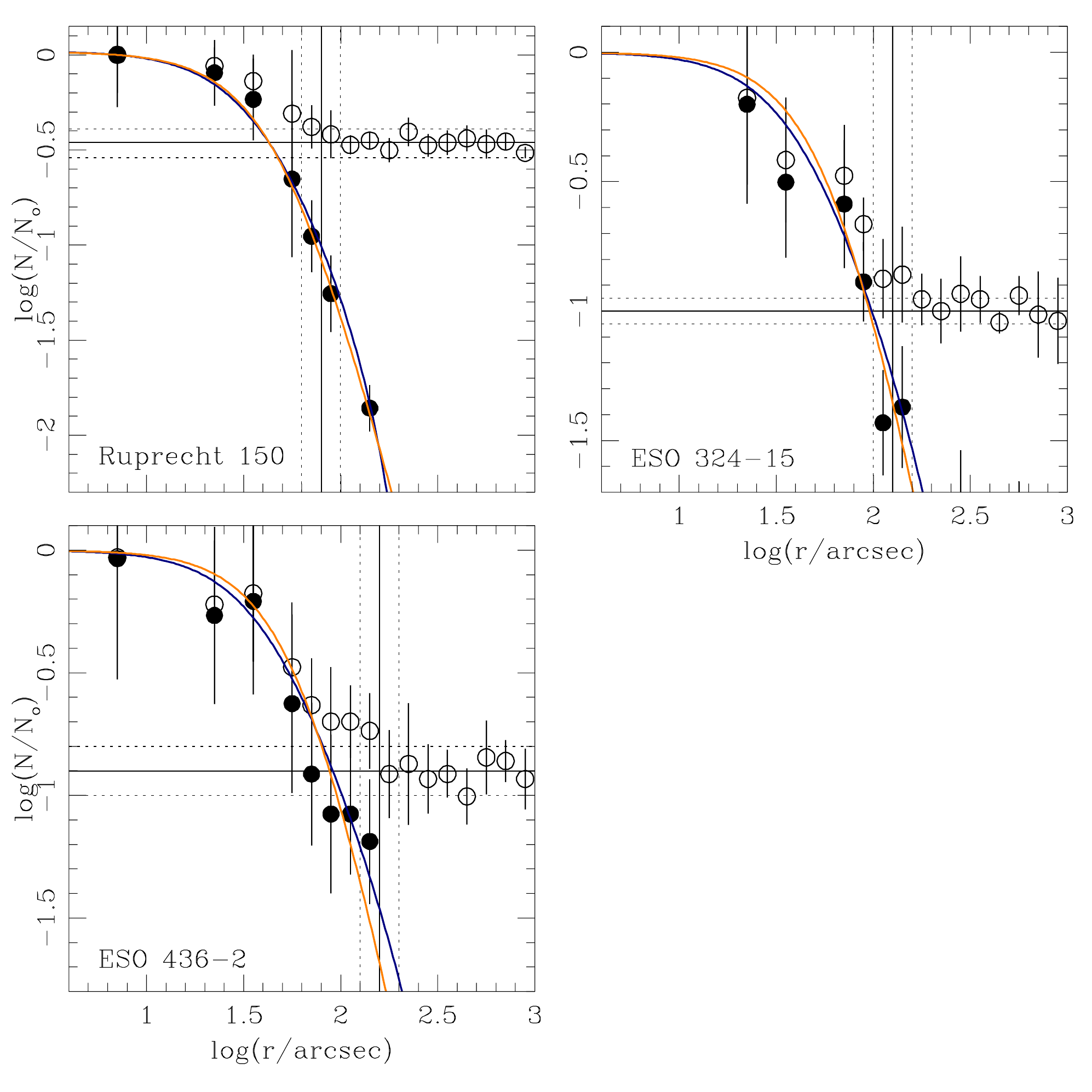}
    \caption{Stellar density profiles normalised to the central density N$_o$ obtained 
from star counts. Open and filled circles 
refer to measured and
background subtracted density profiles, respectively. 
 The background level and the cluster radius are indicated by solid horizontal
and vertical lines, respectively; their uncertainties are in dotted lines.
Blue and orange solid lines
depict the fitted King and Plummer curves, respectively.}
    \label{fig:fig2}
\end{figure*}

The resulting density profiles 
expressed as number of stars per arcsec$^2$ are shown in  Fig.~\ref{fig:fig2}.
In the figure, we represent the constructed and
background subtracted density profiles with open and filled circles, respectively.
Errorbars represent rms errors  of star counts among boxes at the same radius, 
to which we added the mean error of the
background star count to the background subtracted density profile. The background level and the cluster radius are indicated by solid horizontal
and vertical lines, respectively; their uncertainties are in dotted lines.
The  normalised background corrected density profiles  (those of Fig.~\ref{fig:fig2}) were fitted using a \citet{king62} model through the expression 

\begin{equation}
 N \varpropto ({\frac{1}{\sqrt{1+(r/r_c)^2}} - \frac{1}{\sqrt{1 + (r_t/r_c)^2}}})^2
,\end{equation}

\noindent where $r_c$ and $r_t$ are the core and tidal radii, respectively.

We also fitted \citet{plummer11} profiles using the expression

\begin{equation}
N \varpropto \frac{1}{(1+(r/a)^2)^2} 
,\end{equation}

\noindent where $a$ is the Plummer radius, which is related to the half-mass radius ($r_h$) by the relation $r_h$ $\sim$ 1.3$a$.  We used a grid of $r_c$ and $r_t$ and  $r_h$ radii to minimize $\chi$$^2$ while fitting both King and Plummer profiles,
separately. The values which best reproduce
the observed stellar density profiles are listed in Table~\ref{tab:table10} and the
respective King and Plummer curves are  plotted with blue and orange solid lines in 
Fig.~\ref{fig:fig2}, respectively. Their uncertainties were estimated by taking into 
account the dispersion in the fitted density profiles.


\section{CMD cleaning}

We built six colour-magnitude diagrams (CMDs) and three colour-colour (CC)
diagrams by extracting every star from 
our $UBVRI-CT_1T_2$ photometric data sets located within the cluster radii
 ($r_{cls}$). Since they
account for the luminosity function, colour distribution and
stellar density of the stars distributed along the cluster line of sights, 
we first statistically cleaned them before estimating the
cluster fundamental parameters.

We performed such a cleaning of field stars by employing the procedure developed by 
\citet[see their Fig. 12]{pb12} and also used elsewhere 
\citep[e.g.][and references therein]{p14,petal15a,petal15b,pb16}. The method compares a 
extracted cluster CMD to
distinct CMDs composed of stars located reasonably far from the
object, but not too far so as to risk losing the local field-star
signature in terms of stellar density, luminosity function and/or
colour distribution. Here we chose four field regions, each one designed to cover an
equal area as that  of the cluster, and placed around the cluster
 at $\sim$ 3-4$\times$$r_{cls}$ from the cluster centre.
Note that the four selected fields could not
adequately represent the fore/background of the cluster if the extinction varies
significantly accross the field of view. The procedure carries out the comparison 
between field-star and cluster CMDs by using boxes which vary their
sizes from one place to another throughout the CMD and are centred on the
positions of every star found in the field-star CMD.

Since we repeated this task for each of the four field CMD box
samples, we could assign a membership probability to each star in the
cluster CMD. This was done by counting the number of times a star
remained unsubtracted in the four cleaned cluster CMDs and by
subsequently dividing this number by four. Thus, we distinguished
field populations projected on to the cluster area, i.e., those stars
with a probability $P$ $\le$ 25\%; stars that could equally likely be
associated with either the field or the object of interest ($P$ =
50\%); and stars that are predominantly found in the cleaned cluster
CMDs ($P \ge$ 75\%) rather than in the field-star CMDs. 
Statistically speaking, a certain amount of cleaning residuals is expected, 
which depends on the degree of variability of the stellar density,
luminosity function and colour distribution of the field stars.

Figures~\ref{fig:fig3} to \ref{fig:fig9} show the whole set of CMDs and 
CC diagrams for the cluster sample that can be exploited from the present 
extensive multi-band photometry. They include every magnitude and colour 
measurements of stars located within the respective cluster radii 
(see Table~\ref{tab:table10}). We have also incorporated to the figures
the statistical photometric memberships obtained above by
distinguishing stars with different colour symbols as follows: stars that
statistically belong to the field ($P \le$ 25\%, pink), stars that might belong 
to either the field or the cluster ($P =$ 50\%, light blue), and stars that 
predominantly populate the cluster region ($P \ge$ 75\%, dark blue). At first glance, 
the cleaned cluster CMDs (stars with $P \ge$ 75\%) resemble those relatively poor or 
poor -in terms of number of stars- intermediate-age open clusters.

\subsection{Proper motion}

Three clusters (Ruprecht\,9, 37 and 150) have been studied by \citet{diasetal2014}
using data from the UCAC4 \citep{zachariasetal2013} catalog in a homogeneous way.
They derived mean proper motions of the clusters and the membership probabilities of the 
stars in the region of each cluster by applying the statistical method based on a 
global optimization procedure to fit the observed distribution of 
proper motions with two overlapping normal bivariate frequency functions, which also 
take the individual proper motion errors into account. We used here their membership
probabilities to compare them with those coming from our photometric procedure, and
to provide an independent support of identifying the fiducial CMD cluster sequences.

For the remaining clusters in the sample (Ruprecht\,3, 74, ESO\,324-15 and 436-2)
\citet{diasetal2014} could not apply their procedure, because of the small number of stars with
proper motion measurements in the cluster fields. Every star with both photometric 
and proper motion  membership probabilities $P \ge$ 75\% were encircled 
in Figs.~\ref{fig:fig3} to \ref{fig:fig9}. We also searched for parallaxes and proper motions
 measured by the {\it Gaia} satellite \citep{gaia2016} and
any available information was used in Section 6.

\section{Cluster fundamental parameters}

The availability of six CMDs and three different CC diagrams covering wavelengths from
the blue up to the  near-infrared
allowed us to derive reliable ages, reddenings and distances for the studied clusters. 
Cluster fundamental parameters were estimated following by
matching theoretical isochrones to the various CMDs and CC diagrams, simultaneosly.

We used the theoretical isochrones of \citet{betal12} for [Fe/H] from -0.7 
to +0.2 dex, in steps of $\Delta$[Fe/H] = 0.1 dex. The adopted metallicity range covers 
most of that for the well-studied Milky Way open clusters 
\citep[see, e.g.][]{paunzeretal2010,hetal14}. For each [Fe/H] value we made use
of the shape of the main sequence (MS), its curvatures (those  less and more pronounced), 
the relative distance between the giant stars and the main sequence turnoff (MSTO) in magnitude and colour separately, 
among others, to find the age of the isochrone which best matches the cluster's features in the 
CMDs and CC diagrams, regardless the cluster reddening and distance.
From our best choice (this includes both [Fe/H] and age values), we derived the cluster 
reddenings by shifting that isochrone in the three CC diagrams
following the reddening vectors until their bluest points coincided with the observed ones.
Note that this requirement allowed us to use the $V-R$ vs $R-I$ CC diagram as well, even 
though the reddening vector runs almost parallell to the cluster sequence.
Finally, the mean $E(B-V)$ colour excesses were used to properly shift the chosen isochrone
in the CMDs in order to derive the cluster true distance moduli $(m-M)_o$ by shifting the isochrone 
along the magnitude axes. 

In order to enter the isochrones into the CMDs and CC diagrams we 
used the following ratios: $E(U-B)$/$E(B-V)$ = 0.72 + 0.05$\times$$E(B-V)$ \citep{hj56}; 
$E(V-R)$/$E(B-V)$ = 0.65, $E(V-I)$/$E(B-V)$ = 1.25, $A_{V}$/$E(B-V)$ = 3.1 \citep{cetal89}; 
$E(C-T_1)$/$E(B-V)$ = 1.97, $E(T_1-T_2)$/$E(B-V)$ = 0.692, $A_{T_1}$/$E(B-V)$ = 2.62 
\citep{g96}.

We found that isochrones bracketing the age choiced  by $\Delta$
log($t$ yr$^{-1}$) = $\pm$0.10 and $\Delta$[Fe/H] = $\pm$0.10 dex represent the overall 
age/metallicity uncertainties
owing to the observed dispersion in the cluster CMDs and CC diagrams.  Fig.~\ref{fig:fig4}
shows the adopted isochrone (solid line) and two additional ones (dashed and dotted lines)
to illustrate the overall uncertainties.
The adopted best matched isochrones are overplotted on Figs.~\ref{fig:fig3} to \ref{fig:fig9}
with black solid lines, 
while the resulting values with their errors for the cluster reddenings, distances, 
ages and metallicities are listed in Table~\ref{tab:table10}. 

We additionally applied a global optimization fitting method, the Cross-Entropy (CE) technique \citep{monteiroetal2017}, to estimate the fundamental parameters of the studied clusters. This is a completely independent and self-consistent procedure for analizying clusters $UBVRI$ data
sets. As for the CE parameter tuning, we followed the prescriptions outlined in \citet{monteiroetal2017}, except the binary fraction considered here was 50$\%$ 
\citep{hurleyetal2007,letal12}. We used the central coordinates and radii obtained in Section 3 and
the stars with $P \ge$ 75\% (see Section 4). Because of the small number of stars in the CMDs we did not perform neither the variation of the IMF nor the bootstrap technique. Finally, we considered the photometric errors.
We found cluster parameters (reddening, distance, age and
metallicity) in agreement with those derived above within
the quoted uncertainties.

%


\section{results and discussion}


Ruprecht\,3 has been studied by \citet{pavanietal2003} from 2MASS data \citep{skrutskieetal2006}.
They estimated an age of log($t$ yr$^{-1}$) = 9.20$\pm$0.15, a distance from the Sun of 
0.72$\pm$0.04 kpc and a colour excess $E(B-V)$= 0.04 from isochrone fitting assuming solar metal 
content. They suggested that the properties of Ruprecht\,3 are compatible with what would be expected 
for an intermediate-age open cluster remnant, and described the object like a
poorly populated compact concentration. Here we have made use of parallaxes ($\pi$) and proper motions
($\mu$) measured by
the {\it Gaia} satellite \citep{gaia2016} for five bright stars in the cluster field. They have been 
identified in Fig.~\ref{fig:fig3} with the numbers \#1 to 5 and their $\pi$ (mas), 
$\mu$$_{\rm RA}$ (mas/yr), $\mu$$_{\rm DEC}$ (mas/yr) values are (1.02$\pm$0.48, 1.775$\pm$1.680,   
  -0.398$\pm$1.729)$_{\rm 1}$, (1.02$\pm$0.35, -1.027$\pm$1.747, -7.641$\pm$1.494)$_{\rm 2}$,
(1.64$\pm$0.25, -0.129$\pm$1.234, 5.415$\pm$1.071)$_{\rm 3}$, (0.81$\pm$0.46, -6.115$\pm$2.593,
5.366$\pm$0.996)$_{\rm 4}$ and (3.71$\pm$0.31, -13.987$\pm$1.541, 3.563$\pm$1.312), respectively.
 As can be seen, ($\pi$, $\mu$$_{\rm RA}$, $\mu$$_{\rm DEC}$) differ among the stars, a result
that agrees with the different photometric memberships of them, in the sense that values
from {\it Gaia} are not expected to be identical across a sample including both members and 
non-members.  From $\pi$ values we found that the five bright stars are located between
270 up to 1230 pc from the Sun, while their proper motions differ significantly compared
to the known dispersion in stellar aggregates \citep{diasetal2014}.
From these values, we conclude that Ruprecht\,3 is not an open cluster remnant.


\citet{bb10} also took advantange of 2MASS data to derive an age of 3$\pm$1 Gyr, a distance
from the Sun of 5.25$\pm$0.74 kpc and a colour excess $E(B-V)$ = 0.00$\pm$0.06 mag 
for Ruprecht\,37. The cluster age
agrees pretty well with our value. However, its distance differs significantly.
The authors assumed [Fe/H] = 0.0 dex which is quite different to our estimated value [Fe/H] = -0.5 dex,
based on the metallicity sensitive $C-T_1$ and $U-B$ colours. The different metallicity could
affect the estimation of the remaining fundamental parameters, as well as, the fact that the
2MASS photometry is much  shallower than the present one. The former barely reaches the cluster
MSTO, which our photometric data sets go down $\sim$ 3 mag below the MSTO. The left-hand
panel of Fig~\ref{fig:fig10} shows the 2MASS CMD for stars located within the cluster radius.
We have represented with blue filled circles stars that have photometric and proper motions
merbership probabilitiess (see Sect. 4) higher than 75 and 70 per cent, respectively.
As can be seen, the 2MASS data show a important scatter compared to the $J-H$ colour
range, which makes it less reliable even when using bona-fide cluster stars. 

ESO\,324-15 was classified by \citet{pavanietal2011} as a probable physical system, and 
estimated for it an age of log($t$ yr$^{-1}$) = 9.02$\pm$0.12, a colour excess
$E(B-V)$ = 0.13$\pm$0.07 mag and a distance from the Sun of  0.94$\pm$0.15 kpc. They adopted
[Fe/H] = 0.0 dex. Every fundamental parameter is in fairly good agreement with those
derived here, except the cluster distance. The impact of such a difference can be assessed
in the right-hand panel of Fig~\ref{fig:fig10} where we have overplotted their selected isochrone
with a solid red line.

ESO\,436-02 seems to be made up of the brightest stars in the observed CMDs and CC diagrams.
During the isochrone fitting, we considered as MSTO stars those at $V$ $\sim$ 10 mag
(see Fig.~\ref{fig:fig9}). As for the CE solutions, the cluster parameters 
do not show Gaussian distributions as typically happens when dealing
with star clusters, and  the final results are not satisfactory for both
the blue $B-V$ and red $V-I$ colours (see Fig~\ref{fig:fig9}).
If we considered MSTO stars those with $V$ $<$ 12 mag, neither the isochrone fitting nor
the CE method would support such an hypothesis. The {\it Gaia} parallaxes and proper
motions for four bright stars, numbered \#1 to 4 in Fig~\ref{fig:fig9} are 
($\pi$ (mas), $\mu$$_{\rm RA}$ (mas/yr), $\mu$$_{\rm DEC}$ (mas/yr) = 
(2.23$\pm$0.32, -3.509$\pm$0.630, -7.741$\pm$0.392))$_{\rm 1}$, (3.24$\pm$0.28, 
-0.400$\pm$0.674, -2.707$\pm$0.434)$_{\rm 2}$, (3.90$\pm$0.87, -19.779$\pm$2.699,
7.093$\pm$1.018)$_{\rm 3}$ and (3.75$\pm$0.26, -42.952$\pm$0.676, -11.185$\pm$0.402),
respectively. From these values we conclude that the considered stars do not form
a physical system.

\setcounter{table}{9}
\begin{table*}
\centering
\caption{Derived properties of selected open clusters.}
\label{tab:table10}
\begin{tabular}{@{}lccccccccccc}\hline
Star cluster & $E(B-V)$ & $(m-M)_o$ & d  &  $r_c$ & $r_h$ &  $r_{cls}$ &  $r_t$  & $\log(t)$ & [Fe/H] & $t_r$\\
             &  (mag)  & (mag)  & (kpc)    &   (pc) &  (pc) &   (pc)     &  (pc)  &  & (dex)  & (Myr)  \\\hline
Ruprecht\,3  &   0.25$\pm$0.05  & 10.0$\pm$0.1  & 1.00$^{+0.04}_{-0.05}$ &  0.14$\pm$0.02 & 0.31$\pm$0.03 & 0.52$^{+0.02}_{-0.03}$ & 1.70$\pm$0.24 & 9.0 & 0.0 &  $--$ \\
Ruprecht\,9  &   0.05$\pm$0.02  & 12.0$\pm$0.1  & 2.51$^{+0.10}_{-0.11}$ &  0.61$\pm$0.06 & 1.11$\pm$0.08 & 1.10$^{+0.06}_{-0.07}$ & 3.04$\pm$0.61 & 9.4 & 0.0 &  5.67$\pm$1.06\\   
Ruprecht\,37 &   0.20$\pm$0.04  & 12.3$\pm$0.1  & 2.88$^{+0.13}_{-0.14}$ &  0.49$\pm$0.07 & 0.91$\pm$0.09 & 1.26$^{+0.08}_{-0.09}$ & 3.50$\pm$0.70 & 9.6 &-0.5 & 4.10$\pm$0.92 \\
Ruprecht\,74 &   0.00$\pm$0.03  & 11.3$\pm$0.1  & 1.82$^{+0.09}_{-0.11}$ &  0.31$\pm$0.04 & 0.63$\pm$0.06 & 0.90$^{+0.05}_{-0.06}$ & 3.53$\pm$0.44 & 9.3 & 0.0 &   2.49$\pm$0.54\\
Ruprecht\,150&   0.10$\pm$0.05  & 10.5$\pm$0.1  & 1.26$^{+0.05}_{-0.06}$ &  0.21$\pm$0.03 & 0.40$\pm$0.04 & 0.48$^{+0.03}_{-0.04}$ & 1.53$\pm$0.30 & 9.2 & 0.0 &  1.00$\pm$0.29\\
ESO\,324-15  &   0.15$\pm$0.04  &  9.0$\pm$0.1  & 0.63$^{+0.01}_{-0.03}$ &  0.12$\pm$0.01 & 0.26$\pm$0.02 & 0.38$^{+0.02}_{-0.03}$ & 1.38$\pm$0.15 & 9.1 & 0.0 &  0.67$\pm$0.13\\
ESO\,436-2   &   0.00$\pm$0.05  &  8.9$\pm$0.1  & 0.60$^{+0.01}_{-0.02}$ &  0.12$\pm$0.01 & 0.25$\pm$0.02 & 0.46$^{+0.01}_{-0.02}$ & 1.61$\pm$0.15 & 9.1 &-0.2 &  $--$\\
\hline
\end{tabular}
\noindent Note: to convert 1 arcsec to pc, we use the following expression,10$\times$10$^{(m-M)_o/5}$sin(1/3600)
where $(m-M)_o$ is the true distance modulus.

\end{table*}

\setcounter{figure}{9}
\begin{figure}
	\includegraphics[width=\columnwidth]{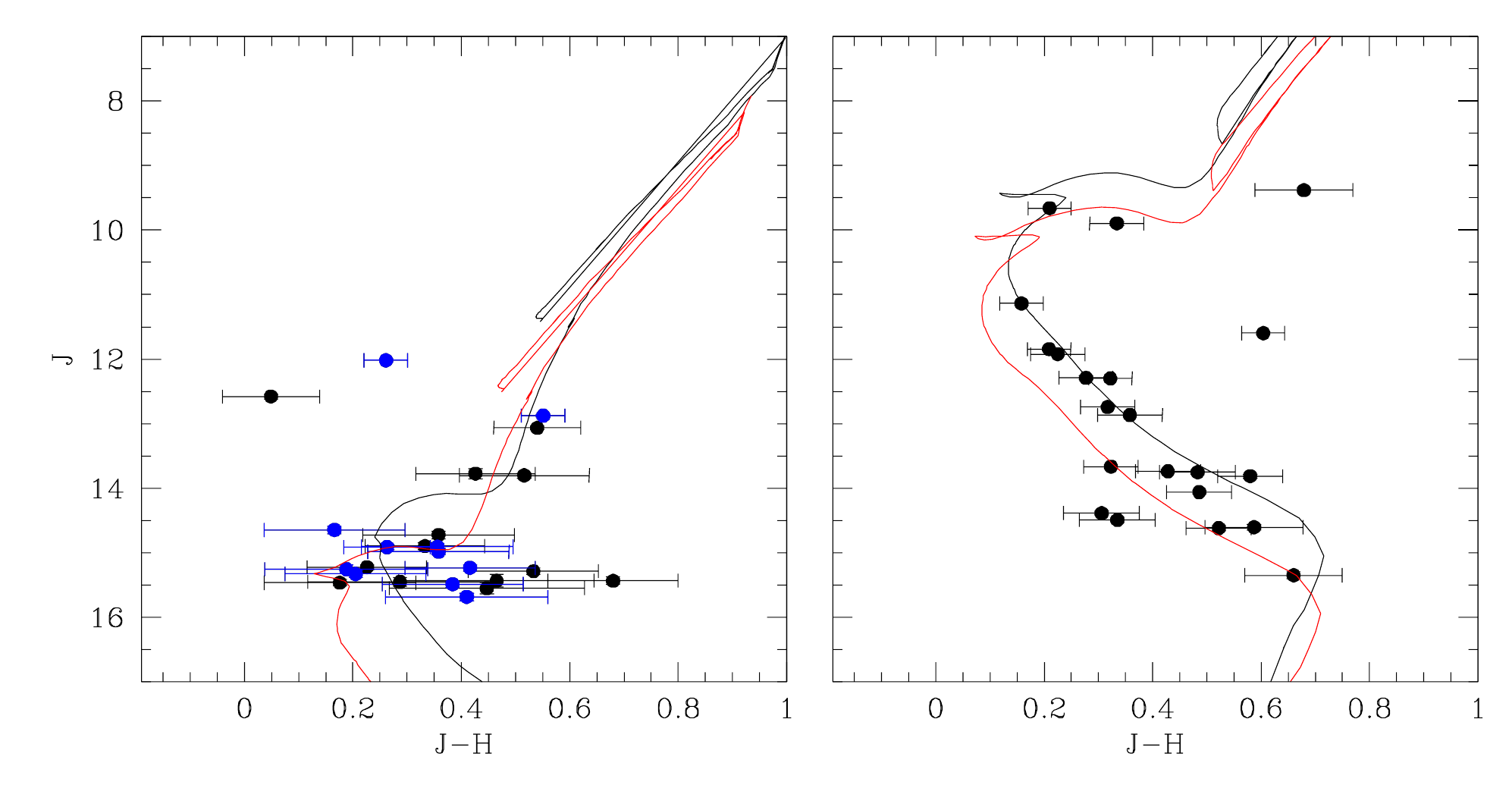}
    \caption{2MASS CMDs for stars located within the cluster radius of Ruprecht\,37 (left-hand panel), 
and ESO\,324-15 (right-hand panel). Blue filled circles (and
respective errorbars) represent
stars with photometric and proper motion membership probabilities higher than 75 and 70 per cent,
respectively. Theoretical isochrones of \citet{betal12} for the cluster fundamental 
parameters estimated previously and in this work are superimposed with solid red and black
lines, respectively (see Sect. 6 for details). }
    \label{fig:fig10}
\end{figure}

\begin{table}
\caption{Derived masses ($\msun$) of selected open clusters.}
\label{tab:table11}
\begin{tabular}{@{}lccc}\hline
Star cluster & $M_{total}$&$M_{J16}$ & $M_{Salpeter}$  \\
 \hline
Ruprecht\,9  & 10.3$\pm$4.7 & 82$\pm$88  &  130  \\   
Ruprecht\,37 & 46.3$\pm$21.3 & 70$\pm$77 &  125 \\
Ruprecht\,74 & 13.8$\pm$6.3 & 90$\pm$96  &  150 \\
Ruprecht\,150& 13.5$\pm$6.2 & 97$\pm$101 &  170 \\
ESO\,324-15  & 14.1$\pm$6.4 & 106$\pm$111&  135 \\
\hline
\end{tabular}
\end{table}

We derived the masses of Ruprecht\,9, 37, 74, 150 and ESO\,324-15 by summing the individual masses of 
stars with membership probabilities $P \ge$ 75\%. The latter were obtained by interpolation 
in the theoretical isochrones traced in Figs.~\ref{fig:fig4} to \ref{fig:fig8} from the 
observed $V$ magnitude of each star, properly corrected by reddening and distance modulus. 
We estimate the uncertainty in the mass to be  $\sigma(\log(M/\msun))$ $\sim$ 0.2 
dex. Note that this error comes from propagation of the $V$ magnitude errors in the
mass distribution along the theoretical isochrones. It does not reflect the
deviation of the cluster mass computed from stars with $P \ge$ 75\%
from the  actual cluster mass. Nevertheless, at first glance, the appearance of 
the cluster CMDs and CC diagrams ($P \ge$ 75\%) do not seem to significantly differ
from those including any other observed stars placed along the adopted isochrones
with $P <$ 75\%, thought to be cluster stars. The derived 
 total cluster masses ($M_{total}$) of Ruprecht\,9, 37,
74, 150 and ESO\,324-15  are listed in Table~\ref{tab:table11}. The resulting cluster mass functions (MFs) are
shown in Fig.~\ref{fig:fig11} where the errorbars come from applying Poisson statistics. 
The solid lines represent the relationship
given by \citet[][slope = -2.35]{salpeter55} for the stars in the solar neighbourhood.

We also used the relationship between the  cluster mass and the cluster age
derived by \citet[][their equation 8]{joshietal2016} from 489 open clusters
located closer than 2 kpc from the Sun, to estimate the masses of the clusters
in our sample. The  computed masses ($M_{J16}$)  are listed in 
 Table~\ref{tab:table11}.
As can be seen, the $M_{total}$ values resulted to be $\sim$ 10-15 per cent  of $M_{J16}$ 
ones,
except for Ruprecht\,37 whose $M_{total}$ is $\sim$ 65 per cent of its  $M_{J16}$ value.
Alternatively,  in order to have some other rough estimate of upper mass limits,
we used the Salpeter's law along with the zero points in Fig.~\ref{fig:fig11}, to estimate the
 cluster mass  ($M_{Salpeter}$) down to a star mass of 0.5$\msun$. We derived  masses 
35-40 per cent
larger  than $M_{J16}$, except for ESO\,324-15
whose  $M_{Salpeter}$ value resulted to be 20 per cent larger that those from \citet{joshietal2016},
respectively  (see Table~\ref{tab:table11}).

Using the half-mass radii $r_h$ and masses of Table~\ref{tab:table11}, we
computed the half-mass relaxation times using the equation \citep{sh71}:

\begin{equation}
t_r = \frac{8.9\times 10^5 M_{cls}^{1/2} r_h^{3/2}}{\bar{m} log_{10}(0.4M_{cls}/\bar{m})}
,\end{equation}

\noindent where $M_{cls}$ is the  $M_{J16}$ value and $\bar{m}$ is the mean mass of 
the cluster  stars. We used $\sigma$$M_{J16}$ and $\sigma$$r_h$ to estimate $\sigma$$t_h$. 
The resulting $t_r$ values are
listed in the last column of Table~\ref{tab:table10}.

\setcounter{figure}{10}
\begin{figure*}
	\includegraphics[width=\textwidth]{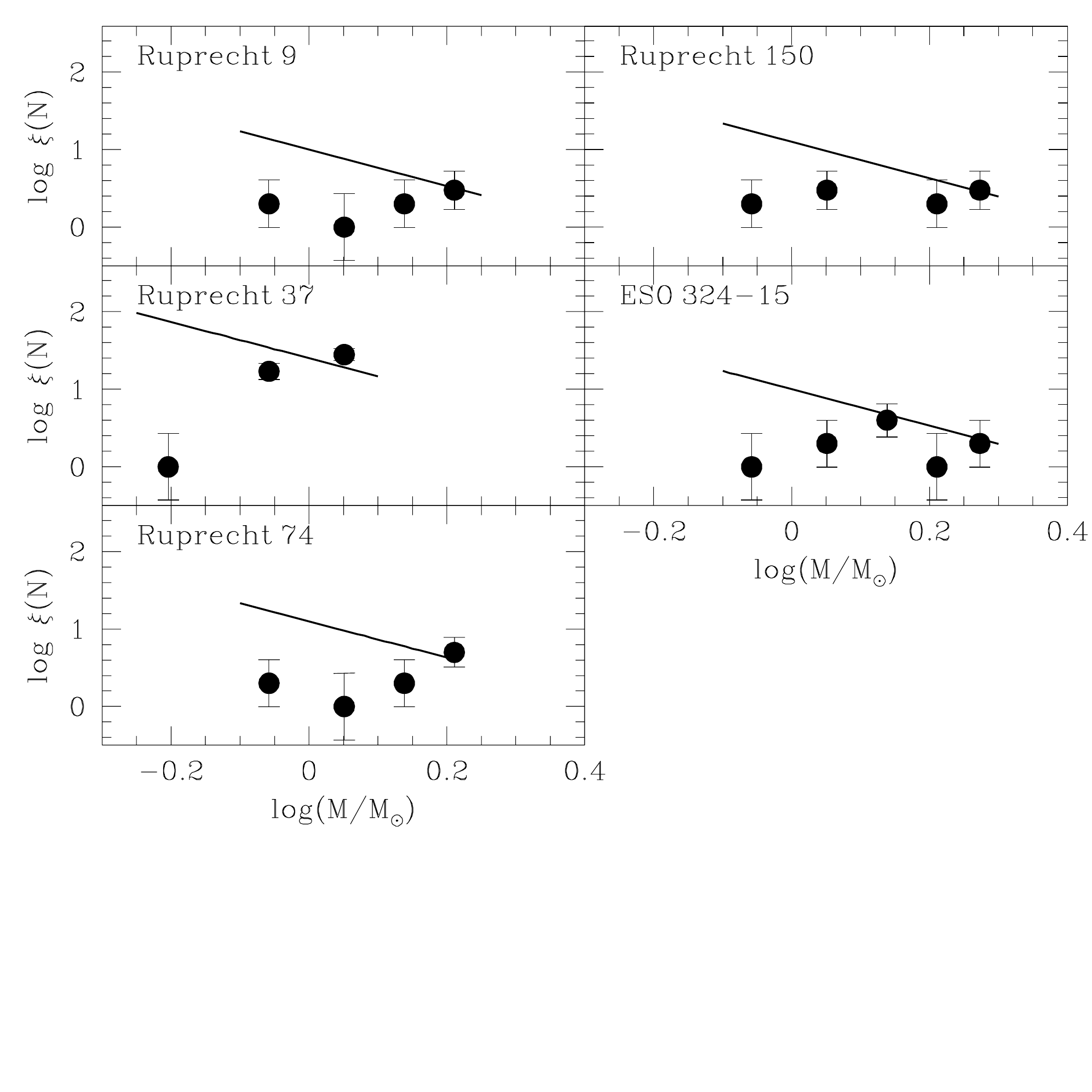}
    \caption{Mass function for clusters in our sample. The solid lines represent 
the relationship given by \citet[][slope = -2.35]{salpeter55} for the stars in the 
solar neighbourhood.}
    \label{fig:fig11}
\end{figure*}

The comparison of the cluster ages with their respective $t_r$ values clearly reveals that all 
the clusters in the sample have survived hundreds of times their characteristic two-body relaxation times 
and should be much closer to a disruption stage. Their  total masses, which represent a mass loss around 
85-90 per cent from their $M_{J16}$ values -with the sole exception of Ruprecht\,37 whose mass loss is around 
35 per cent-, also confirm such a highly evolved dynamical stage.  If we considered the 
$M_{Salpeter}$ values as the initial cluster masses instead, the mass loss would be still larger.
Note that the mass loss is due to both
internal dynamical evolution and tidal effects. As for the Galactic tidal field, 
\citet[][see, e.g., their figures 1]{miholicsetal2014} found that the difference in the potential  wall 
between 6 and 100 kpc from 
the galactic centre leads to $\le$ 10 per cent variations in the half-mass radius for clusters younger than
4 Gyr. Thus, bearing in mind the Galactocentric distances of the studied clusters ($R_{GC}$ = 9 $\pm$ 1 kpc) 
and their ages (see Table~\ref{tab:table10}), we considered the dynamical evolution as the main
origin of mass loss. The effect of an important mass loss 
stand out in Fig~\ref{fig:fig11} where the observed MFs depart from that of the Salpeter's law 
towards lower stellar masses. Surprinsingly enough is the fact that even though the surviving clusters 
keep small amounts of their  $M_{J16}$ and $M_{Salpeter}$  masses, the cluster CMDs are still useful to derive their fundamental 
paramaters.

From the analysis of the derived structural paramaters, it is also feasible to draw conclusions about
their present dynamical stage. \citet{trentietal2010} presented a unified picture for the evolution of star clusters 
on the two-body relaxation timescale from direct N-body simulations of star clusters
in a tidal field. Their treatment of the stellar evolution is based on the approximation
that most of the relevant stellar evolution occurs on a timescale shorter than a 
relaxation time, when the most massive stars lose a significant fraction of mass and consequently  
contribute to a global expansion of the system. Later in the
life of a star cluster, two-body relaxation tends to erase the memory of the initial 
density profile and concentration. They found that the
structure of the system, as measured by the core to half-mass radius ratio, the
concentration parameter $c$= log($r_t/r_c$), among others, 
evolve toward a universal state, which is set by the efficiency of heating on
the visible population of stars induced by dynamical interactions in the core of
the system. The concentration parameter $c$
for this model increases steadily with time.

The resulting $c$ values for our clusters are within 0.9 and 1.1 (0.7 for Ruprecht\,9). 
These values correspond to star clusters in an advance stage of dynamical evolution.
Indeed, we compared our $c$ values with those for 236 open clusters analyzed by \citet{piskunovetal2007}, 
who derived from them homogeneous scales of radii and masses. They derived core and tidal radii for 
their cluster sample, from which we calculated the half-mass radii and, with their clusters masses and 
equation 10, relaxation times, by assuming that the cluster stellar density
profiles can be indistinguishably reproduced by King and Plummer models.
Their cluster sample is mostly distributed inside 
a circle of $\sim$ 1 kpc from the Sun and has $c$ values between $\sim$ 0.1 up to 1.1 
following a broad trend with the  age/$t_r$ ratio, in the sense that the larger 
the $c$ values, the more dynamically evolved an open cluster.

According to \citet[][see, e.g., their figure 33.2]{hh03} a star cluster dynamically
evolving with its tidal radius filled, moves in the $r_c/r_h$ vs $r_h/r_t$ plane 
parallel to the $r_c/r_h$ axis ($r_h/r_t$ $\sim$ 0.21) toward low values due to 
violent relaxation in the cluster core region followed by two-body relaxation,
mass segregation, and finally core-collapse. The derived $r_h/r_t$ ratio for the present
cluster sample is 0.22 $\pm$ 0.07 (see Table~\ref{tab:table10}), which is in excellent agreement 
with the expected value for a tidally filled cluster. Curiously, the tidal radii are quit
similar to the Jacobi radii, which confirm that the studied clusters are tidally filled. 
The latter were calculated using the expresion \citep{cw90} :

\begin{equation}
r_J = (\frac{M_{cls}}{3 M_{gal}})^{1/3}\times R_{GC},
\end{equation}

\noindent where $M_{cls}$ is the cluster mass and $M_{gal}$ is the Milky Way (MW) mass 
inside the cluster galactocentric distance $R_{GC}$. To compute $M_{gal}$  (= 5$\times$10$^{11}$\msun)
we used the MW mass
profile of \citet{tayloretal2016}. We found that $|$$r_t$-$r_J$$|$ $<$ $\sigma$$r_t$ + $\sigma$$r_J$,
meaning that cluster stars have occupied as much as possible the allowed volume without being 
stripped away from the cluster. 
In the same model by \citet{hh03}, the $r_c/r_h$ ratio ranges from 
1.4 (start of evolution) down to 0.1. The  $r_c/r_h$ ratio of the studied clusters is 0.49 $\pm$ 0.06, 
which confirms their evolved dynamical stages.

\section{conclusions}

We present a comprehensive multi-band photometric analysis of 
seven catalogued open clusters, namely: Ruprecht\,3, 9, 37, 74, 150, ESO\,324-15 and 436-2.
The objects were observed 
through the Johnson $UBV$,  Kron-Cousins $RI$ and Washington $C$ filters;  four of 
them (Ruprecht\,9, 74, 150 and ESO\,436-2) are photometrically study for the first time, while
for Ruprecht\,3, 37 and ESO\,324-15, our photometric data sets surpass that from 2MASS photometry.

The multi-band photometric data sets were used to trace the cluster stellar
density radial profiles and to build CMDs and CC diagrams, from which we estimated their
structural parameters and fundamental astrophysical properties. Cluster radii were
derived from a careful placement of the background levels in the
radial profiles  built from star count throughout the observed
fields using the final photometric catalogues. We fitted King and Plummer models to 
derive cluster core, half-mass and tidal radii. 

The constructed cluster CMDs and CC diagrams were statistically cleaned from field star 
contamination using a powerful technique that makes use of cells varying in position and size
in order to reproduce the field CMD as closely as possible. Then, from six cleaned CMDs and three
cleaned CC diagrams covering  wavelengths from the blue up to the  near-infrared we
estimated the cluster fundamental parameters. We exploited such a wealth in combination with theoretical
isochrones to find out that the clusters in our sample are of intermediate-age (9.0 $\le$ log($t$ yr$^{-1}$) 
$\le$ 9.6), of relatively small size ($r_{cls}$ $\sim$ 0.4 $-$ 1.3 pc) and placed between 0.6 and 2.9 kpc
from the Sun. Their  total masses, computed by summing the individual masses of stars with photometric
membership probabilities $P \ge$ 75\%, resulted to be $\sim$ 10-15 per cent of the  cluster
masses estimated from an independent robust calibration of the cluster mass as a function of
the cluster age. The cluster MFs built using the same sample of stars also account for so high 
percentage of mass loss. We found that Ruprecht\,3 and ESO\,436-2 do not show self-consistent
evidence to be physical systems.

We compared the cluster masses, concentration parameters,  $r_c/r_h$, $r_h/r_t$ and 
age/$t_r$ ratios to those for 236 clusters located in the solar neighbourhood as well as
to different theoretical models. We conclude that the studied  clusters should be 
much closer to
their disruption stage as a result of their internal dynamical evolution (mass segregation)
and Galactic tidal effects. The stars with photometric membership probabilities $P \ge$ 75\%
occupy a volume as large as those for tidally filled clusters.

\section*{Acknowledgements}
This work has made use of data from the European Space Agency (ESA)
mission {\it Gaia} (\url{http://www.cosmos.esa.int/gaia}), processed by
the {\it Gaia} Data Processing and Analysis Consortium (DPAC,
\url{http://www.cosmos.esa.int/web/gaia/dpac/consortium}). Funding
for the DPAC has been provided by national institutions, in particular
the institutions participating in the {\it Gaia} Multilateral Agreement.
We thank the anonymous referee whose thorough comments and suggestions
allowed us to improve the manuscript.




\bibliographystyle{mnras}

\begin{thebibliography}{}
\makeatletter
\relax
\def\mn@urlcharsother{\let\do\@makeother \do\$\do\&\do\#\do\^\do\_\do\%\do\~}
\def\mn@doi{\begingroup\mn@urlcharsother \@ifnextchar [ {\mn@doi@}
  {\mn@doi@[]}}
\def\mn@doi@[#1]#2{\def\@tempa{#1}\ifx\@tempa\@empty \href
  {http://dx.doi.org/#2} {doi:#2}\else \href {http://dx.doi.org/#2} {#1}\fi
  \endgroup}
\def\mn@eprint#1#2{\mn@eprint@#1:#2::\@nil}
\def\mn@eprint@arXiv#1{\href {http://arxiv.org/abs/#1} {{\tt arXiv:#1}}}
\def\mn@eprint@dblp#1{\href {http://dblp.uni-trier.de/rec/bibtex/#1.xml}
  {dblp:#1}}
\def\mn@eprint@#1:#2:#3:#4\@nil{\def\@tempa {#1}\def\@tempb {#2}\def\@tempc
  {#3}\ifx \@tempc \@empty \let \@tempc \@tempb \let \@tempb \@tempa \fi \ifx
  \@tempb \@empty \def\@tempb {arXiv}\fi \@ifundefined
  {mn@eprint@\@tempb}{\@tempb:\@tempc}{\expandafter \expandafter \csname
  mn@eprint@\@tempb\endcsname \expandafter{\@tempc}}}

\bibitem[\protect\citeauthoryear{{Bonatto} \& {Bica}}{{Bonatto} \&
  {Bica}}{2010}]{bb10}
{Bonatto} C.,  {Bica} E.,  2010, \mn@doi [\mnras]
  {10.1111/j.1365-2966.2010.17000.x}, \href
  {http://adsabs.harvard.edu/abs/2010MNRAS.407.1728B} {407, 1728}

\bibitem[\protect\citeauthoryear{{Bressan}, {Marigo}, {Girardi}, {Salasnich},
  {Dal Cero}, {Rubele}  \& {Nanni}}{{Bressan} et~al.}{2012}]{betal12}
{Bressan} A.,  {Marigo} P.,  {Girardi} L.,  {Salasnich} B.,  {Dal Cero} C.,
  {Rubele} S.,   {Nanni} A.,  2012, \mn@doi [\mnras]
  {10.1111/j.1365-2966.2012.21948.x}, 427, 127

\bibitem[\protect\citeauthoryear{{Cardelli}, {Clayton}  \& {Mathis}}{{Cardelli}
  et~al.}{1989}]{cetal89}
{Cardelli} J.~A.,  {Clayton} G.~C.,   {Mathis} J.~S.,  1989, \mn@doi [\apj]
  {10.1086/167900}, 345, 245

\bibitem[\protect\citeauthoryear{{Chernoff} \& {Weinberg}}{{Chernoff} \&
  {Weinberg}}{1990}]{cw90}
{Chernoff} D.~F.,  {Weinberg} M.~D.,  1990, \mn@doi [\apj] {10.1086/168451},
  351, 121

\bibitem[\protect\citeauthoryear{{Dias}, {Alessi}, {Moitinho}  \&
  {L{\'e}pine}}{{Dias} et~al.}{2002}]{detal02}
{Dias} W.~S.,  {Alessi} B.~S.,  {Moitinho} A.,   {L{\'e}pine} J.~R.~D.,  2002,
  \mn@doi [\aap] {10.1051/0004-6361:20020668}, 389, 871

\bibitem[\protect\citeauthoryear{{Dias}, {Monteiro}, {Caetano}, {L{\'e}pine},
  {Assafin}  \& {Oliveira}}{{Dias} et~al.}{2014}]{diasetal2014}
{Dias} W.~S.,  {Monteiro} H.,  {Caetano} T.~C.,  {L{\'e}pine} J.~R.~D.,
  {Assafin} M.,   {Oliveira} A.~F.,  2014, \mn@doi [\aap]
  {10.1051/0004-6361/201323226}, \href
  {http://adsabs.harvard.edu/abs/2014A%26A...564A..79D} {564, A79}

\bibitem[\protect\citeauthoryear{{Gaia Collaboration}, {Brown}, {Vallenari},
  {Prusti}, {de Bruijne}, {Mignard}, {Drimmel}  \& {co-authors}}{{Gaia
  Collaboration} et~al.}{2016}]{gaia2016}
{Gaia Collaboration} {Brown} A.~G.~A.,  {Vallenari} A.,  {Prusti} T.,  {de
  Bruijne} J.,  {Mignard} F.,  {Drimmel} R.,   {co-authors} .,  2016, preprint,
  \href {http://adsabs.harvard.edu/abs/2016arXiv160904172G} {} (\mn@eprint
  {arXiv} {1609.04172})

\bibitem[\protect\citeauthoryear{{Geisler}}{{Geisler}}{1996}]{g96}
{Geisler} D.,  1996, \mn@doi [\aj] {10.1086/117799}, 111, 480

\bibitem[\protect\citeauthoryear{{Heggie} \& {Hut}}{{Heggie} \&
  {Hut}}{2003}]{hh03}
{Heggie} D.,  {Hut} P.,  2003, {The Gravitational Million-Body Problem: A
  Multidisciplinary Approach to Star Cluster Dynamics}

\bibitem[\protect\citeauthoryear{{Heiter}, {Soubiran}, {Netopil}  \&
  {Paunzen}}{{Heiter} et~al.}{2014}]{hetal14}
{Heiter} U.,  {Soubiran} C.,  {Netopil} M.,   {Paunzen} E.,  2014, \mn@doi
  [\aap] {10.1051/0004-6361/201322559}, 561, A93

\bibitem[\protect\citeauthoryear{{Hiltner} \& {Johnson}}{{Hiltner} \&
  {Johnson}}{1956}]{hj56}
{Hiltner} W.~A.,  {Johnson} H.~L.,  1956, \mn@doi [\apj] {10.1086/146231},
  \href {http://adsabs.harvard.edu/abs/1956ApJ...124..367H} {124, 367}

\bibitem[\protect\citeauthoryear{{Hurley}, {Aarseth}  \& {Shara}}{{Hurley}
  et~al.}{2007}]{hurleyetal2007}
{Hurley} J.~R.,  {Aarseth} S.~J.,   {Shara} M.~M.,  2007, \mn@doi [\apj]
  {10.1086/517879}, \href {http://adsabs.harvard.edu/abs/2007ApJ...665..707H}
  {665, 707}

\bibitem[\protect\citeauthoryear{{Joshi}, {Dambis}, {Pandey}  \&
  {Joshi}}{{Joshi} et~al.}{2016}]{joshietal2016}
{Joshi} Y.~C.,  {Dambis} A.,  {Pandey} A.~K.,   {Joshi} S.,  2016, preprint,
  \href {http://adsabs.harvard.edu/abs/2016arXiv160606425J} {} (\mn@eprint
  {arXiv} {1606.06425})

\bibitem[\protect\citeauthoryear{{King}}{{King}}{1962}]{king62}
{King} I.,  1962, \mn@doi [\aj] {10.1086/108756}, 67, 471

\bibitem[\protect\citeauthoryear{{Landolt}}{{Landolt}}{1992}]{l92}
{Landolt} A.~U.,  1992, \mn@doi [\aj] {10.1086/116242}, \href
  {http://adsabs.harvard.edu/abs/1992AJ....104..340L} {104, 340}

\bibitem[\protect\citeauthoryear{{Li}, {Mao}, {Chen}  \& {Zhang}}{{Li}
  et~al.}{2012}]{letal12}
{Li} Z.,  {Mao} C.,  {Chen} L.,   {Zhang} Q.,  2012, \mn@doi [\apjl]
  {10.1088/2041-8205/761/2/L22}, 761, L22

\bibitem[\protect\citeauthoryear{{Miholics}, {Webb}  \& {Sills}}{{Miholics}
  et~al.}{2014}]{miholicsetal2014}
{Miholics} M.,  {Webb} J.~J.,   {Sills} A.,  2014, \mn@doi [\mnras]
  {10.1093/mnras/stu1951}, \href
  {http://adsabs.harvard.edu/abs/2014MNRAS.445.2872M} {445, 2872}

\bibitem[\protect\citeauthoryear{{Monteiro}, {Dias}, {Hickel}  \&
  {Caetano}}{{Monteiro} et~al.}{2017}]{monteiroetal2017}
{Monteiro} H.,  {Dias} W.~S.,  {Hickel} G.~R.,   {Caetano} T.~C.,  2017,
  \mn@doi [\na] {10.1016/j.newast.2016.08.001}, \href
  {http://adsabs.harvard.edu/abs/2017NewA...51...15M} {51, 15}

\bibitem[\protect\citeauthoryear{{Paunzen}, {Heiter}, {Netopil}  \&
  {Soubiran}}{{Paunzen} et~al.}{2010}]{paunzeretal2010}
{Paunzen} E.,  {Heiter} U.,  {Netopil} M.,   {Soubiran} C.,  2010, \mn@doi
  [\aap] {10.1051/0004-6361/201014131}, \href
  {http://adsabs.harvard.edu/abs/2010A%26A...517A..32P} {517, A32}

\bibitem[\protect\citeauthoryear{{Pavani}, {Bica}, {Ahumada}  \&
  {Clari{\'a}}}{{Pavani} et~al.}{2003}]{pavanietal2003}
{Pavani} D.~B.,  {Bica} E.,  {Ahumada} A.~V.,   {Clari{\'a}} J.~J.,  2003,
  \mn@doi [\aap] {10.1051/0004-6361:20021920}, \href
  {http://adsabs.harvard.edu/abs/2003A%26A...399..113P} {399, 113}

\bibitem[\protect\citeauthoryear{{Pavani}, {Kerber}, {Bica}  \&
  {Maciel}}{{Pavani} et~al.}{2011}]{pavanietal2011}
{Pavani} D.~B.,  {Kerber} L.~O.,  {Bica} E.,   {Maciel} W.~J.,  2011, \mn@doi
  [\mnras] {10.1111/j.1365-2966.2010.17999.x}, \href
  {http://adsabs.harvard.edu/abs/2011MNRAS.412.1611P} {412, 1611}

\bibitem[\protect\citeauthoryear{{Piatti}}{{Piatti}}{2014}]{p14}
{Piatti} A.~E.,  2014, \mn@doi [\mnras] {10.1093/mnras/stu534}, 440, 3091

\bibitem[\protect\citeauthoryear{{Piatti}}{{Piatti}}{2016}]{piatti16b}
{Piatti} A.~E.,  2016, \mn@doi [\mnras] {10.1093/mnras/stw2248}, \href
  {http://adsabs.harvard.edu/abs/2016MNRAS.tmp.1355P} {}

\bibitem[\protect\citeauthoryear{{Piatti} \& {Bastian}}{{Piatti} \&
  {Bastian}}{2016}]{pb16}
{Piatti} A.~E.,  {Bastian} N.,  2016, preprint, \href
  {http://adsabs.harvard.edu/abs/2016arXiv160306891P} {} (\mn@eprint {arXiv}
  {1603.06891})

\bibitem[\protect\citeauthoryear{{Piatti} \& {Bica}}{{Piatti} \&
  {Bica}}{2012}]{pb12}
{Piatti} A.~E.,  {Bica} E.,  2012, \mn@doi [\mnras]
  {10.1111/j.1365-2966.2012.21694.x}, 425, 3085

\bibitem[\protect\citeauthoryear{{Piatti}, {de Grijs}, {Rubele}, {Cioni},
  {Ripepi}  \& {Kerber}}{{Piatti} et~al.}{2015a}]{petal15a}
{Piatti} A.~E.,  {de Grijs} R.,  {Rubele} S.,  {Cioni} M.-R.~L.,  {Ripepi} V.,
   {Kerber} L.,  2015a, \mn@doi [\mnras] {10.1093/mnras/stv635}, 450, 552

\bibitem[\protect\citeauthoryear{{Piatti} et~al.,}{{Piatti}
  et~al.}{2015b}]{petal15b}
{Piatti} A.~E.,  et~al., 2015b, \mn@doi [\mnras] {10.1093/mnras/stv2054}, \href
  {http://adsabs.harvard.edu/abs/2015MNRAS.454..839P} {454, 839}

\bibitem[\protect\citeauthoryear{{Pijloo}, {Portegies Zwart}, {Alexander},
  {Gieles}, {Larsen}, {Groot}  \& {Devecchi}}{{Pijloo}
  et~al.}{2015}]{pijlooetal15}
{Pijloo} J.~T.,  {Portegies Zwart} S.~F.,  {Alexander} P.~E.~R.,  {Gieles} M.,
  {Larsen} S.~S.,  {Groot} P.~J.,   {Devecchi} B.,  2015, \mn@doi [\mnras]
  {10.1093/mnras/stv1546}, 453, 605

\bibitem[\protect\citeauthoryear{{Piskunov}, {Schilbach}, {Kharchenko},
  {R{\"o}ser}  \& {Scholz}}{{Piskunov} et~al.}{2007}]{piskunovetal2007}
{Piskunov} A.~E.,  {Schilbach} E.,  {Kharchenko} N.~V.,  {R{\"o}ser} S.,
  {Scholz} R.-D.,  2007, \mn@doi [\aap] {10.1051/0004-6361:20077073}, \href
  {http://adsabs.harvard.edu/abs/2007A%26A...468..151P} {468, 151}

\bibitem[\protect\citeauthoryear{{Plummer}}{{Plummer}}{1911}]{plummer11}
{Plummer} H.~C.,  1911, \mn@doi [\mnras] {10.1093/mnras/71.5.460}, \href
  {http://adsabs.harvard.edu/abs/1911MNRAS..71..460P} {71, 460}

\bibitem[\protect\citeauthoryear{{Rossi}, {Bekki}  \& {Hurley}}{{Rossi}
  et~al.}{2016}]{rossietal2016}
{Rossi} L.~J.,  {Bekki} K.,   {Hurley} J.~R.,  2016, \mn@doi [\mnras]
  {10.1093/mnras/stw1827}, \href
  {http://adsabs.harvard.edu/abs/2016MNRAS.462.2861R} {462, 2861}

\bibitem[\protect\citeauthoryear{{Salpeter}}{{Salpeter}}{1955}]{salpeter55}
{Salpeter} E.~E.,  1955, \mn@doi [\apj] {10.1086/145971}, \href
  {http://adsabs.harvard.edu/abs/1955ApJ...121..161S} {121, 161}

\bibitem[\protect\citeauthoryear{{Skrutskie} et~al.,}{{Skrutskie}
  et~al.}{2006}]{skrutskieetal2006}
{Skrutskie} M.~F.,  et~al., 2006, \mn@doi [\aj] {10.1086/498708}, \href
  {http://adsabs.harvard.edu/abs/2006AJ....131.1163S} {131, 1163}

\bibitem[\protect\citeauthoryear{{Spitzer} \& {Hart}}{{Spitzer} \&
  {Hart}}{1971}]{sh71}
{Spitzer} Jr. L.,  {Hart} M.~H.,  1971, \mn@doi [\apj] {10.1086/150855}, 164,
  399

\bibitem[\protect\citeauthoryear{{Stetson}, {Davis}  \& {Crabtree}}{{Stetson}
  et~al.}{1990}]{setal90}
{Stetson} P.~B.,  {Davis} L.~E.,   {Crabtree} D.~R.,  1990, in {Jacoby} G.~H.,
  ed.,  Astronomical Society of the Pacific Conference Series Vol. 8, CCDs in
  astronomy. pp 289--304

\bibitem[\protect\citeauthoryear{{Taylor}, {Boylan-Kolchin}, {Torrey},
  {Vogelsberger}  \& {Hernquist}}{{Taylor} et~al.}{2016}]{tayloretal2016}
{Taylor} C.,  {Boylan-Kolchin} M.,  {Torrey} P.,  {Vogelsberger} M.,
  {Hernquist} L.,  2016, \mn@doi [\mnras] {10.1093/mnras/stw1522}, \href
  {http://adsabs.harvard.edu/abs/2016MNRAS.461.3483T} {461, 3483}

\bibitem[\protect\citeauthoryear{{Trenti}, {Vesperini}  \& {Pasquato}}{{Trenti}
  et~al.}{2010}]{trentietal2010}
{Trenti} M.,  {Vesperini} E.,   {Pasquato} M.,  2010, \mn@doi [\apj]
  {10.1088/0004-637X/708/2/1598}, \href
  {http://adsabs.harvard.edu/abs/2010ApJ...708.1598T} {708, 1598}

\bibitem[\protect\citeauthoryear{{Zacharias}, {Finch}, {Girard}, {Henden},
  {Bartlett}, {Monet}  \& {Zacharias}}{{Zacharias}
  et~al.}{2013}]{zachariasetal2013}
{Zacharias} N.,  {Finch} C.~T.,  {Girard} T.~M.,  {Henden} A.,  {Bartlett}
  J.~L.,  {Monet} D.~G.,   {Zacharias} M.~I.,  2013, \mn@doi [\aj]
  {10.1088/0004-6256/145/2/44}, \href
  {http://adsabs.harvard.edu/abs/2013AJ....145...44Z} {145, 44}

\makeatother
\end{thebibliography}

\input{paper.bbl}


\setcounter{figure}{2}
\begin{landscape}
\begin{figure} 
\caption{CMDs and CC diagrams for stars measured in the field of Rupreht\,3.
Colour-scaled symbols represent stars with photometric memberships 
$P \le$ 25\% (pink), equals to 50\% (light blue) and $\ge$ 75\% (dark
blue), respectively. We overplotted the isochrones which best matches the
cluster features (black solid line). Stars with {\it Gaia} parallaxes and
proper motions are numbered from \#1 to 5 (see text for details).}
   \label{fig:fig3}
\end{figure}
\end{landscape}

\begin{landscape}
\begin{figure} 
\caption{CMDs and CC diagrams for stars measured in the field of Ruprecht\,9. Symbols
are as in Fig.~\ref{fig:fig3}. Dashed and dotted lines correspond to isochrones
for (log($t$ yr$^{-1}$), [Fe/H]) = (9.3,-0.1) and (9.5,+0.1), respectively.
Large open circles represent stars with both 
photometric ($P \ge$ 75\%) and proper motion ($P \ge$ 75\%) membership probabilities
(see text for details).}
   \label{fig:fig4}
\end{figure}
\end{landscape}

\begin{landscape}
\begin{figure} 
\caption{CMDs and CC diagrams for stars measured in the field of Ruprecht\,37. Symbols
are as in Fig.~\ref{fig:fig3}.   Large open circles represent stars with both 
photometric ($P \ge$ 75\%) and proper motion ($P \ge$ 75\%) membership probabilities
(see text for details).}
\label{fig:fig5}
\end{figure}
\end{landscape}

\begin{landscape}
\begin{figure} 
\caption{CMDs and CC diagrams for stars measured in the field of Ruprecht\,74. Symbols
are as in Fig.~\ref{fig:fig3}.}
\label{fig:fig6}
\end{figure}
\end{landscape}

\begin{landscape}
\begin{figure} 
\caption{CMDs and CC diagrams for stars measured in the field of Ruprecht\,150. Symbols
are as in Fig.~\ref{fig:fig3}.  Large open circles represent stars with both 
photometric ($P \ge$ 75\%) and proper motion ($P \ge$ 75\%) membership probabilities
(see text for details).} 
\label{fig:fig7}
\end{figure}
\end{landscape}

\begin{landscape}
\begin{figure} 
\caption{CMDs and CC diagrams for stars measured in the field of ESO\,324-15. Symbols
are as in Fig.~\ref{fig:fig3}. }
\label{fig:fig8}
\end{figure}
\end{landscape}

\begin{landscape}
\begin{figure} 
\caption{CMDs and CC diagrams for stars measured in the field of ESO\,436-2. Symbols
are as in Fig.~\ref{fig:fig3}. Stars with {\it Gaia} parallaxes and
proper motions are numbered from \#1 to 4. (see text for details).}
\label{fig:fig9}
\end{figure}
\end{landscape}





\appendix


\bsp	
\label{lastpage}


\end{document}